# Studies of Particle Acceleration by an Active Microwave Medium

Paul Schoessow[1], Alexei Kanareykin[1], Levi Schächter[2], Yuriy Bogachev[3], Andrey Tyukhtin[4], Elena Bagryanskaya[5], Natalia Yevlampieva[4]

[1] *Euclid Techlabs LLC, Solon, OH 44139, USA*
[2] *Technion - Israel Institute of Technology, Haifa 32000, Israel*
[3] *St. Petersburg Electrical Engineering University, St. Petersburg 197376, Russia*
[4] *St.Petersburg University, St.Petersburg, Russia, 198302*
[5] *International Tomography Center SB RAS, Novosibirsk, Russia, 630090*

**Abstract.** The PASER is potentially a very attractive method for particle acceleration, in which energy from an active medium is transferred to a charged particle beam. The effect is similar to the action of a maser or laser with the stimulated emission of radiation being produced by the virtual photons in the electromagnetic field of the beam. We have been investigating the possibility of developing a demonstration PASER operating at X-band. The less restrictive beam transport and device dimensional tolerances required for working at X-band rather than optical frequencies as well as the widespread application of X-band hardware in accelerator technology all contribute to the attractiveness of performing a PASER demonstration experiment in this frequency range. Key to this approach is the availability of a new class of active materials that exhibit photoinduced electron spin polarization. We will report on the status of active material development and measurements, numerical simulations, and progress towards a planned microwave PASER acceleration experiment at the Argonne Wakefield Accelerator facility.

**Keywords:** paser, maser, active paramagnetic media, ESR, EPR, fullerene, $C_{60}$

**PACS:** 29.17.+w, 41.75.Jv, 84.40.Ik, 41.75.Lx

## INTRODUCTION

The original PASER concept [1-4] focused principally on the use of an optical or infrared wavelength active medium to provide the energy for accelerating electrons. Recent studies of electron paramagnetic resonance (EPR) in nematic liquid crystal solutions of fullerene or porphyrin compounds have demonstrated activity (negative imaginary part of the magnetic susceptibility) in the ~10 GHz frequency range when the material is optically pumped [5,6]. Unlike conventional solid state maser materials, the $C_{60}$ based materials can operate at relatively high temperatures. Building on prior work in this area we have developed and synthesized new active paramagnetic materials with improved gain and bandwidth properties over those previously reported in the literature and performed CW and time-resolved electron paramagnetic resonance measurements demonstrating the activity of the media.

In order to evaluate the usefulness of the active media for particle acceleration, we are building a test system to measure the electromagnetic properties of active media loaded prototype accelerating structures using standard microwave techniques. We are also evaluating options for a demonstration accelerator experiment.

Besides the EPR measurements and the experimental design efforts that are the main subject of this paper, we are also proceeding with simulation and theoretical activities in support of the microwave PASER project. Refs. [14-16] provide some additional details.

## PROGRESS IN ACTIVE PARAMAGNETIC MATERIALS; $C_{60}$-LC AS A HIGH TEMPERATURE MASER MEDIUM

The first optically pumped solid-state maser based on electron paramagnetic resonance (EPR) was proposed in 1957 [7] and demonstrated successfully in 1962 [8]. To achieve amplification in a paramagnetic material based maser, one must achieve an inverted spin population. These original masers did not find wide applications as among other limitations they require low temperatures ~10 K for operation. A recently published article [0], however, reports the observation of maser amplification at near room temperature in an optically pumped fullerene-based material at ≈9 GHz— an ideal frequency match to the beam parameters currently available at accelerator test facilities such as AWA. Three-level solid-state maser amplifiers based upon the paramagnetic properties of the unpaired electron exploit changes in the electron spin population of the magnetic Zeeman levels, thus allowing amplification of electromagnetic radiation in the bulk material [11].

The $C_{60}$ fullerene molecule was first identified in 1985 [9] as a third stable form of elemental carbon. Unlike diamond and graphite, $C_{60}$ is soluble in a large number of organic solvents. The active X-band media that we have been studying are primarily solutions of $C_{60}$ in liquid crystals. While there are still many aspects of the physics of the maser process in these materials that are not yet completely understood, the general picture is clear.

This work relies heavily on both CW and time-resolved EPR (electron paramagnetic resonance, also known as ESR, electron spin resonance) measurements [10] of candidate LC-fullerene materials. The basic EPR resonance condition in a paramagnetic spin system is $\hbar\omega_0 = g\beta H_0$, where $\omega_0$ is the resonance frequency, $\beta$ is the Bohr magneton, and H is the magnetic field strength. A free electron has g=2.002322. The g-factor varies for the most part between 1 and ~2 depending on spin-orbit interactions in the material; many radicals and paramagnetic complexes have g-factors larger than 2.

Typically EPR measurements are made with a constant frequency X-band microwave source, and the spectrum is actually obtained by sweeping the applied magnetic field (H~3000 Oe for λ~3 cm). (This will have interesting consequences for the use of these materials in accelerators—tuning the resonant frequency can be accomplished by changing the magnetic field.) Finally, the gyromagnetic ratio is defined as $\gamma = g\beta/\hbar$.

The population difference $\Delta N$ between the spin levels is directly related to $\chi''$ (the imaginary part of the magnetic susceptibility) which is indirectly measured in EPR experiments. The $C_{60}$ molecule in its ground state is not EPR active. However after photoexcitation of $C_{60}$ by a short pump laser or flash lamp pulse an EPR active triplet state can be formed. In ordinary solvents like toluene or benzene, the dipole-dipole interaction averages to zero; there is no zero field splitting of the triplet state. The unsplit state can only absorb rf energy at $\omega_0 = \gamma H_0$.

The EPR spectra and activity of $C_{60}$ are strongly dependent on the solvent used. It is important for our purposes that $C_{60}$ in different organic solvents or liquid crystals is light sensitive and undergoes reversible changes in the optical or EPR absorption spectra. We have synthesized and studied a number of $C_{60}$ based materials to study the effect of solvent type, $C_{60}$ concentration, and additional solutes like porphyrins on the activity of the media. One particular material, a $C_{60}$–porphyrin-LC solution was found to exhibit significantly improved amplification properties over previously studied substances.

The effect of the nematic liquid crystal component (rodlike molecules that align to exhibit a long range 1D order) is to introduce a symmetry breaking. The dipole-dipole interaction in the triplet state does not completely average to zero). The triplet state in each fullerene molecule now undergoes a Zeeman splitting into three levels. If the spin energy levels of the fullerene triplet state become selectively populated, stimulated emission can occur.

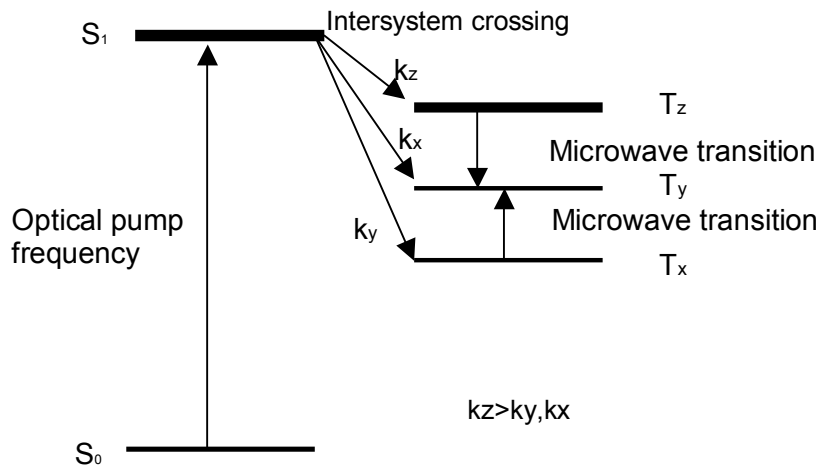

**FIGURE 1**. Energy level diagram of the active $C_{60}$-LC medium. Transitions to $T_y$ can be either absorptive or emissive, as observed in our measurements.

The activity in these materials originates in chemically induced dynamic electron spin polarization (CIDEP). CIDEP is formed in reactions of radicals resulting from optical pumping that in turn lead to the formation of free radicals or triplet state molecules. There are several different mechanisms of CIDEP– triplet [13], radical pair, radical pair-triplet etc.

Figure 1 shows a schematic energy level diagram for the active media discussed in this report. After absorption of the laser pulse molecules transition from the $S_0$ (initial singlet) state to the $S_1$ (excited) state. Following a fast intersystem crossing the excited molecule transitions from $S_1$ to the triplet $T_1$ state. In the triplet state the spin levels are split into three levels (in the case of negligible hyperfine interaction with nuclei). As a rule the rate constants of the transitions from the S1 state to the Tx, Ty, Tz states are different resulting in electron polarization due to the triplet mechanism. In the case of a long triplet molecule lifetime the polarization of this triplet molecule can be observed. In many cases the triplet molecule produces radicals (due to chemical reactions) and polarization on these radicals can be observed. Other mechanisms of CIDEP result from interaction of radicals and their spin dynamics and chemical kinetics.

## STORED ENERGY, EFFICIENCY AND OTHER CHALLENGES IN FULLERENE-BASED ACTIVE MEDIA

The stored energy in the active material can be estimated by taking the product of the spin density in the inversion and the energy of the maser transition, or roughly $\Delta n \times h\nu \sim (10^{17} cm^{-3})(6.6 \times 10^{-27} erg \cdot s)(10^{10} Hz) = 6.6\, erg\, cm^{-3}$. Further assuming about 1 $cm^3$ of active material per linear $cm$ of accelerator, we get an accelerating gradient of 660 $keV/cm$ assuming 1 $pC$ of charge in the bunch, 100% energy transfer efficiency, and no dielectric or wakefield losses in the medium. Similarly, assuming a spin relaxation time $\sim 1\mu s$, the power density of the medium is $\sim 0.66\, W/cm^3$.

It is important to note that there is no fundamental reason why the spin density (and consequently the stored energy) cannot be increased further. The currently attainable spin density is measured in solutions where the fullerene concentration is $\sim 10^{-3}\, mol/l$. At larger concentrations there is a tendency for the $C_{60}$ molecules to form clusters; we are currently investigating alternative choices of solvent and preparation technique to ameliorate this.

Also affecting the performance of $C_{60}$–LC media are the relatively large dielectric losses (tan δ>0.005) from the liquid crystal component. Cooling the material to -20 C is probably necessary to keep losses small. So far it seems that large tan δ is intrinsic to the nematic LCs currently available, although reducing the loss tangent by appropriate choice of solvent is not ruled out. Our approach to this problem is to design accelerator and bench test structures that make use of mode symmetries to locate the active medium in a region where the electric field is small and the magnetic field (which couples to the permeability) is large.

We have considered a number of possible implementations of active media devices for particle acceleration. The first involves acceleration of a single bunch by the active medium (the basic PASER concept) in a resonant structure, similar to the dielectric wakefield accelerator, with the fundamental resonant frequency of the structure adjusted to correspond to the frequency of the masing transition. (The medium is a viscous liquid but can be isolated from the beam channel by a thin walled quartz capillary.) Extensive calculations have shown that this design is marginal for studying acceleration for lossy media, since for TM modes there is no region where large B and small E fields can be obtained simultaneously.

A much more promising approach is to use an iris loaded structure partially filled with an active medium and tuned to the frequency of the maser line. In this case the vanishing of the radial electric field on the irises results in a region for approximately $r > 0.7R$ where the magnetic field dominates. For the bench test a single pillbox cavity will be used to obtain the same effect.

The liquid crystal component of the active material has a permittivity ~2 and dielectric loss tangent >0.005. This implies two kinds of loss mechanisms for the PASER that originate in the dielectric, ohmic (Joule heating) and non-dissipative energy losses from Cherenkov radiation/wakefields. While these are nontrivial concerns, the estimated wakefield losses for an 10 GHz device/1 pC beam are roughly ~0.5 keV/m compared to a limiting accelerating gradient of 6.6 MV/m. Dielectric losses are a more serious problem. One option that we are investigating is the possibility of structure modifications so that the active medium is located near a minimum in the electric field but in a region where the magnetic field is sufficiently large. We also are planning to evaluate the use of other liquid crystals. Part of this investigation will involve tests of reduced loss LC materials.

The efficiency of a laser pumped active microwave medium will be relatively small. In the simplest estimate the efficiency is $(\hbar\omega_{microwave})/(\hbar\omega_{pump}) \sim 0.002\%$. This is somewhat offset by the less stringent quality requirements on the optical pump source; clearly, however some form of energy recovery must be used for a practical PASER accelerator based on this medium.

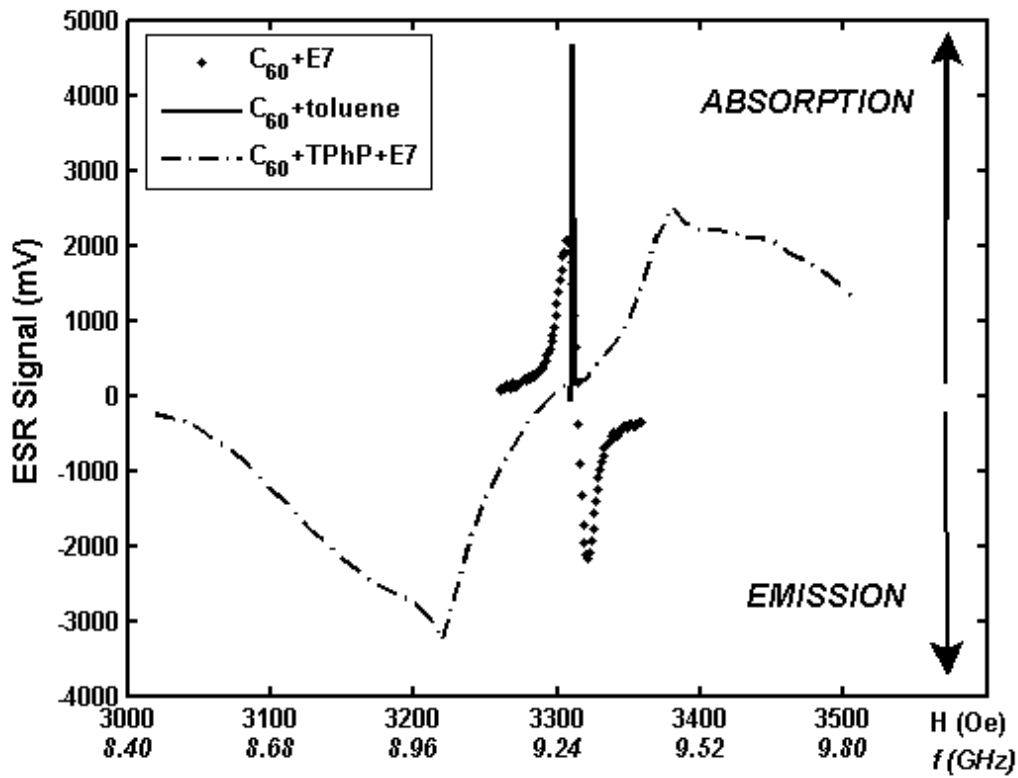

**FIGURE 2.** Comparison of time resolved EPR spectra of active and reference samples at 480ns after laser flash. The $C_{60}$+TPhP+E7 sample clearly shows the highest level of activity (EPR emission signal) as well as the largest bandwidth. (Dotted line: $C_{60}$+E7, t=750 ns, T=243K. Broken line: $C_{60}$+E7+TPhP, t=480 ns, T= 243K. Solid line: $C_{60}$+toluene, t=1.7 µs, T=263 K.)

# EPR MEASURMENTS OF $C_{60}$ –BASED ACTIVE MEDIA

In time resolved (TR) EPR we apply a pump laser pulse to the test sample in the spectrometer. The resulting EPR signal is measured using a digital oscilloscope to obtain a time-resolved spectrum. This technique is especially relevant to the diagnosis of activity in our samples since the generation of a population inversion in the fullerene-liquid crystal solution requires an optical pulse excitation. TR EPR measurements of photoexcited $C_{60}$ in different matrices were carried out using a Bruker ESP-380 CW spectrometer with its field modulation disconnected. The time resolution of the system is 40 ns. The temperature in the EPR resonator was maintained by using a variable temperature nitrogen flow Dewar system. The samples in the microwave cavity of the spectrometer were photoexcited at 532 nm by the second harmonic of a Nd:YAG laser producing 10-ns duration pulses of 15 mJ/pulse at a repetition rate of 10 Hz.

We used this technique to evaluate the activities of different reference and active media samples. We report the results here for $C_{60}$–toluene, $C_{60}$–E7, and $C_{60}$–TPhP–E7. The EPR spectra for these materials are shown in Figure 2.

The toluene solution is used as a reference sample for the active media measurements. After laser flash irradiation of a solution of $C_{60}$ in toluene at 253 K a single absorption EPR line is observed at g = 2.0026. It exhibits a nearly Lorentzian lineshape with width $\Delta B_{1/2}$ = 0.5 G [12]. The presence of only a single absorption peak is due to the fast rotation of the molecules which tends to average the zero field splitting (ZFS) factor to zero.

It is evident that the spectra shown for both the liquid crystal solutions exhibit ESP, demonstrating both emission and absorptive characteristics. (The liquid crystal used as a solvent in these samples is E7, produced commercially by Merck.)

Photoexcitation of a 1:1 mixture of $C_{60}$ and $H_2$Tetraphenylporphyrin ($H_2$TPhP) results in the appearance of a spectrum consisting of two components. A narrow spectrum (~10 G) (the light "break" or "bend" near the zero crossing in Fig. 2) is superimposed on a broad one (~170 G). The broad spectrum may be attributed to the triplet $^3H_2$TPhP (dominant) [16], and the narrow part to $^3C_{60}$.

In this experiment both $C_{60}$ and $H_2$TPhP absorb light. The main polarization is observed on a triplet state of $H_2$TPhP while the polarization of $C_{60}$ is about one order of magnitude less. The spectrum obtained is qualitatively different from that of $C_{60}$–E7. The broad spectrum shows the inversion of absorption – emission lines from low to high magnetic field, respectively. The observed decay time of the polarization is about 1500 ns. The improvement in bandwidth and spin population (by factors of ~10 and ~1.5 respectively) over the $C_{60}$-E7 material is clearly visible in Figure 2.

Despite the attribution of the two components of the spectrum to (predominantly) TPhP and $C_{60}$ the overall form of the spectrum is not simply the sum of the individual spectra. Rather, the spectrum results from the complex interaction of the excited states of the two components in the solution. A detailed analysis of the mechanism of ESP in the $^3C_{60}$-TPhP complex in liquid crystals as well as the possibility of increasing ESP intensity in similar systems is in progress.

We have found that among all the investigated samples that $H_2$TPhP–$C_{60}$ in liquid crystal E7 is the most promising agent for applications in microwave amplification.

The first advantage of this system is the width of the EPR spectrum of the triplet molecule of $H_2TPhP$–$C_{60}$ is 300 MHz for both emission and absorption lines, allowing broad band microwave operation. In addition the reversibility of the photochemical reactions in this material implies the ability to use the medium over many pump-discharge cycles.

## MICROWAVE PASER EXPERIMENTS

At present all of the measurements of active microwave media discussed in this paper have been performed using very small samples in EPR spectrometers. We are building a system to study prototype microwave PASER structures with standard electronic test instruments (Figure 3).

An elliptical cell fabricated from aluminum or other nonferromagnetic material serves both as support for the test cell and flash lamp, while its interior mirrored surface concentrates the light from the flashlamp onto the test cell. The cell dimensions allow it to be mounted inside the bore of the solenoid that provides the magnetic field for the paramagnetic resonance. The flashlamp and test cell are mounted at the foci of the ellipse. Endcaps on the cell will provide mechanical support for the test cavity and flashlamp. The test cavity is an X-band resonator and waveguide that contains the active medium. Measurements of the reflection and transmission coefficients require rf energy to be coupled into and out of the cell. The outer quartz tube is covered with axial conducting strips forming transparent longitudinal slit windows. The conducting strips will be in electrical contact with the endcaps. The strips may be deposited using printed circuit technology; alternatively copper tape or enamel insulated wires could be used. The pumping radiation is input to the medium through these slits. This technique preserves the symmetry of the TM modes.

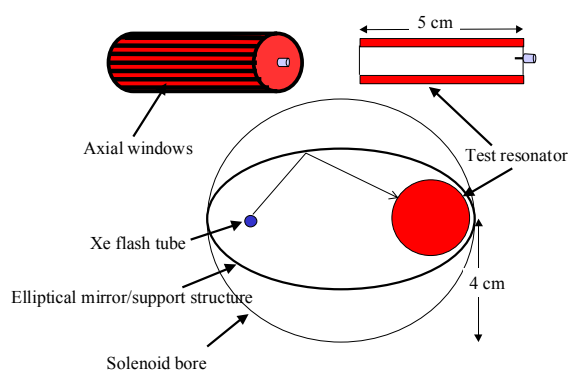

**FIGURE 1.** Sketch of bench test system under construction at the Argonne Wakefield Accelerator lab.

By adjusting the solenoid field we can adjust the resonant frequency of the active medium to coincide with the resonant frequency of each mode of the structure, thus any or all of the X-band modes can be employed for these measurements.

A microwave PASER acceleration demonstration is challenging on a number of levels. The portions of the accelerating structure containing the active medium must be supplied with pump energy from a flashlamp or laser and also maintained in a uniform magnetic field of about 3 kG for X-band operation. We have been considering several possible geometries for PASER experiments. A possible approach is to use a trigger beam to excite a wakefield that is then amplified by the active medium and used to accelerate a second beam [4, 17]. Technologies based on the wakefield amplifier approach are currently favored as the basis for a beam experiment.

One challenge is providing the magnetic field for the medium along the length of the structure, an approach that would require a very long solenoid to implement along the lines of the bench test. We are considering a combination of the following approaches to simplify the design of the beam test structure:

- Permanent magnets for providing the EPR magnetic field. The use of permanent magnets offers an attractive alternative to the use of electromagnets in terms of cost, compactness, and efficiency. Fields in the required range (~3 kG) can be easily obtained. A planar geometry for the medium could be used to minimize the gap size in the magnet and hence the amount of permanent magnet material required.
- Use of fiber optics or light guides to transport the pump radiation to the active material. This allows considerably more flexibility in the configuration of the structure and magnets. In particular the pump signals could be supplied through small gaps in the magnets rather than requiring the structure and light source to be immersed in the magnetic field.
- Nonaxisymmetric structures. Removing the active medium from the beam path results in a considerable simplification of beam transport and reduction of magnet costs (due to the reduction in the magnetic field volume required).

## CONCLUSIONS

A number of chemical systems, consisting of $C_{60}$ and porphyrin molecules in a liquid crystal matrix have been synthesized and EPR tested as candidate active materials for use in X-band microwave PASER experiments. The performance of accelerators based on this technology has been evaluated. A design for a bench test of a microwave PASER cavity prototype has been developed and is under construction at ANL. Some possible options for a beam acceleration experiment have been considered.

## ACKNOWLEDGMENTS

This work was supported by the US Department of Energy, Division of High Energy Physics Grant # DE-FG02-05ER84355 and by the Russian Foundation for Basic Research, RFBR Grant # 06-02-16442-a.